\documentclass{article}
\usepackage[]{psfig}
\newcommand{\be}{\begin{equation}}
\newcommand{\ee}{\end{equation}}
\newcommand{\ba}{\begin{eqnarray}}
\newcommand{\ea}{\end{eqnarray}}
\newcommand{\hp}{\,\hat{\!\bar \phi}}
\begin{document}
\title{\bf Nielsen-Olesen vortices in noncommutative space}
\author{G.S.~Lozano\thanks{Associated with CONICET}\,,\\
{\normalsize\it Departamento de F\'\i sica, FCEyN, Universidad de
Buenos Aires}\\ {\normalsize\it Pab.1, Ciudad Universitaria,
Buenos Aires,Argentina}\\ ~
\\
E.F.~Moreno$^*$ \, and \,
F.A.~Schaposnik\thanks{Associated with CICBA}
\\
{\normalsize\it Departamento de F\'\i sica, Universidad Nacional
de La Plata}\\ {\normalsize\it C.C. 67, 1900 La Plata, Argentina}}

\maketitle

\begin{abstract}
We construct an exact regular vortex solution to the self-dual
equations of the Abelian Higgs model in  non-commutative space for
arbitrary values of $\theta$. To this end, we propose an ansatz
which is the analogous, in Fock space, to the one leading to exact
solutions for the Nielsen-Olesen vortex in commutative space. We
compute the flux and energy of the solution and discuss its
relevant properties.
\end{abstract}
\date{}

%



\section{Introduction}
The recent interest aroused by quantum field theories in
noncommutative space \cite{CDS}-\cite{SW}  prompted the search of
localized classical solutions in noncommutative geometry.
Instantons, solitons carrying various kinds of fluxes, BPS and
non-BPS solutions to different noncommutative theories have been
presented in refs.\cite{HA}-\cite{BLP}. Among these models, the
Abelian-Higgs model in non-commutative space has received
particular attention in connection with vortex like solutions
\cite{GN}-\cite{BLP}.

Several vortex solutions that have been discussed up to now  are
regular at finite noncommutative parameter $\theta$, but they
become singular in the limit $\theta \to 0$ \cite{GN},\cite{JKL}.
More precisely, the magnetic field $B$ associated to the flux tube
behaves as $B \to \delta^{(2)}(x)$ as $\theta \to 0$.
This fact is not surprising since these solutions are obtained
through a procedure which is the analogous to performing singular
gauge transformations leading to topologically non-trivial
solutions from trivial ones in commutative space.

A different class of vortices in non commutative space has been
considered by D.P.Jatkar, G.Mandal and S.Wadia \cite{JMW}, which
are closer in spirit to the regular Nielsen-Olesen  vortices of
the theory in ordinary space. More specifically, in \cite{JMW},
{\it self-dual} Bogomol'nyi equations were derived and the
solutions in the limiting cases $\theta \to 0$  (where they are
regular) and $\theta \to \infty$ were considered.

Bogomol'nyi equations for the Abelian Higgs model were also
discussed in \cite{Bk} and \cite{BLP}. The case considered there
corresponds, in our terminology, to the {\it anti-self-dual case}.
In fact, as in the model in commutative space, there are two sets
of Bogomol'nyi equations, one admitting solutions with positive
flux (which we call self-dual), another  admitting solutions for
negative flux (which we call anti-self-dual). Nevertheless, while
these two sets of equations (and their solutions) are trivially
related via a parity transformation in commutative space, the
presence of the parity breaking parameter $\theta$ prevents such a
trivial connection in the non-commutative case. Hence the
existence and properties of solutions should be checked
separately. While this has been done in detail for the
anti-self-dual case in \cite{Bk} and \cite{BLP}, less is known for
the self-dual case with the exception of the limiting cases
$\theta \to 0$, $\theta \to\infty$.

We present in this note an ansatz leading to regular vortex
solutions in noncommutative space for arbitrary value of $\theta$.
Starting from the noncommutative Abelian Higgs Lagrangian , we
solve the associated self-dual equations  finding an exact
solution that is the noncommutative version of the exact one
presented long ago for the ordinary Nielsen-Olesen vortices
\cite{dVS} and, remarkably, it shares, qualitatively, all  its
basic properties. In fact the solution converges to it as $\theta$
go to zero while, for large values of $\theta$, its profile
differs appreciably from the Nielsen-Olesen solution.

~

We consider  space-time with coordinates $X^\mu$ ($\mu = 0,1,2,3$)
obeying the following noncommutative relations
\be
[X^\mu,X^\nu] = i \theta^{\mu\nu} \label{00}
\ee
We take $\theta^{0i} = 0$ ($i=1,2,3$). Concerning $\theta^{ij}$,
it can be can be brought into its canonical (Darboux) form by an
appropriate orthogonal rotation

\be
[X^1,X^2] = i\theta \, , \;\;\;\;\; [X^1,X^3] = [X^2,X^3] = 0
\label{1} \ee

One way to describe field theories in noncommutative space is by
introducing a Moyal product $*$ between ordinary functions. To
this end, one can establish a one to one correspondence between
operators $\hat f$ and ordinary functions $f$ through a Weyl
ordering
\be
\hat f(X^1,X^2) = \frac{1}{2\pi} \int d^2k \tilde f(k_1,k_2)
\exp\left( i(k_1X^1 + k_2X^2) \right)
\ee
Then, the product of two Weyl ordered operators $\hat f \hat g$
corresponds to a function $f*g(x)$ defined as

\be
f*g(x) =
\left.\exp\left(\frac{i\theta}{2}(\partial_{x_1}\partial_{y_2} -
\partial_{x_2}\partial_{y_1}) \right) f(x_1,x_2) g(y_1,y_2) \right
\vert_{x_1=x_2, y_1 = y_2} \ee

Given a $U(1)$ gauge field $A_\mu(x)$, the field strength
$F_{\mu\nu}$ is defined as
\be
F_{\mu\nu} = \partial_\mu A_\nu - \partial_\nu  A_\mu - i(A_\mu *
A_\nu - A_\nu * A_\mu) \ee
We shall couple the gauge field to a complex scalar field $\phi$
with covariant derivative
\be
D_\mu \phi = \partial_\mu \phi -i A_\mu * \phi \ee
Dynamics for the model will be governed by the Lagrangian
\be
L = -\frac{1}{4} F_{\mu\nu}* F^{\mu \nu}+ \overline{D_\mu \phi}*
D^\mu\phi - \frac{1}{2}( \phi* \bar \phi - \eta^2)^2] \label{6}
\ee
Here we have chosen coefficient of the symmetry breaking potential
at the Bogomol'ny point \cite{dVS}-\cite{Bogo}. We are looking for
static  axially symmetric Nielsen-Olesen vortices with $A_0 = A_3
= 0$. Then, the only relevant coordinates in the problem will be
$i=1,2$.

The alternative approach to noncommutative field theories is to
directly work with operators in the phase space $(X^1,X^2)$, with
commutator (\ref{1}). In this case the $*$ product is just the
product of operators and integration over the $(X^1,X^2)$ plane is
a trace,
\be
\int dx^1dx^2 f(x^1,x^2) = 2\pi \theta {\rm Tr} \hat f(X^1,X^2)
\ee
In this framework, we introduce
complex variables $z$ and $\bar z$
\be
z = \frac{1}{\sqrt{2}}(x^1 + i x^2)\, , \;\;\;\;\;\;
\bar z= \frac{1}{\sqrt{2}}(x^1 - i x^2)
\label{2}
\ee
and annihilation and creation operators  $\hat a$ and $\hat a^\dagger$
in the form
\be
\hat a = \frac{1}{\sqrt{2\theta}}(X^1 + i X^2)\, , \;\;\;\;\;\;
\hat a^\dagger = \frac{1}{\sqrt{2\theta}}(X^1 - i X^2)
\label{3}
\ee
so that (\ref{1}) becomes
\be
[\hat a,\hat a^\dagger] = 1
\label{4}
\ee
With this conventions, derivatives are given by
\be
\partial_z = -\frac{1}{\sqrt \theta}
[\hat a^\dagger,~] \, , \;\;\;\;\;\; \partial_{\bar z} =
\frac{1}{\sqrt \theta} [\hat a,~] \label{5} \ee
The field strength takes then the form
\be
\hat F_{z \bar z} = \partial_z \hat A_{\bar z} - \partial _{\bar
z} \hat A_z -i [\hat A_z,\hat A_{\bar z}] = -
\frac{1}{\sqrt \theta} \left ([\hat a^\dagger,\hat A_z] +
[\hat a,A_{\bar z}] + i \sqrt \theta[\hat A_z,\hat A_{\bar z}]
\right) \equiv i \hat B
\ee
with $\hat B$ the magnetic field. Concerning covariant derivatives
\begin{eqnarray}
D_{\bar z}\hat \phi &=& \partial_{\bar z} \hat \phi - i A_{\bar z}
\hat \phi = \frac{1}{\sqrt \theta} [\hat a,\hat \phi] - i \hat
A_{\bar z} \hat\phi \nonumber\\ D_{z}\hp &=& \partial_z \hp+ i
\hat A_{z} \hp = -\frac{1}{\sqrt \theta} [\hat a^\dagger,\hp] + i
\hat A_z \hp
\end{eqnarray}
where
\be
\hat A_z = \frac{1}{\sqrt {2}} (\hat A_1 - i \hat A_2) \, ,
\;\;\;\; \hat A_{\bar z} = \frac{1}{\sqrt {2}} (\hat A_1 + i \hat
A_2) \ee

The energy functional associated to action (\ref{6}) can be then
written as \cite{JMW}
\be
E = 2\pi \theta {\rm Tr} \left(\frac{1}{2}\hat B^2 + D_{\bar
z}\hat \phi D_{z}\hp +  D_{z}\hat \phi D_{\bar z}\hp +
\frac{1}{2}(\hat\phi \hp - \eta^2)^2 \right) \label{en} \ee

We want to find static solutions minimizing the energy. To this
end, we shall proceed \`a la Bogomol'nyi   writing the energy $E$
in the two following forms \cite{JMW}
\be
E = 2\pi \theta {\rm Tr} \left(\frac{1}{2}\left(\hat B + (\hat\phi
\hp -\eta^2)\right) ^2 + 2D_{\bar z}\hat \phi D_{z}\hp  + (\hat
T^s + \eta^2 \hat B) \right) \label{enb} \ee
with $\hat T^s$ defined as
\be
\hat T^s =  \partial_z ((D_{\bar z} \hat \phi) \hp)
-
\partial_{\bar z} ((D_{z} \hat \phi) \hp)
\ee
or
\be
E = 2\pi\theta {\rm Tr} \left((\frac{1}{2}\left(\hat B - (\hat\phi
\hp -\eta^2)\right) ^2 + 2D_{\bar z} \hp D_{z} \hat\phi  - (\hat
T^a + \eta^2\hat B) \right) \label{enb-} \ee
with
\be
\hat T^a = - \hat T^s \ee Now, one can easily see that ${\rm Tr}
\hat T^a = 0$ \cite{JMW} and hence the energy is bounded by the
magnetic flux, as in the case of  vortices in ordinary space. The
bound is attained when the following first order Bogomol'nyi eqs.
hold
\be
\hat B =    \eta^2 - \hat \phi \hp  \, , \;\;\;\;\; D_{\bar z}
\hat \phi = 0 \, \;\;\;\;\;{\rm self\!\!-\!dual~equations}
\label{beg} \ee
or
\be
-\hat B =   \eta^2 - \hat \phi \hp  \, , \;\;\;\;\; D_z\hat \phi
= 0 \, \;\;\;\;\;{\rm anti~self\!\!-\!dual~equations} \label{beg9}
\ee
We have fixed  in eqs.(\ref{beg})-(\ref{beg9}) our terminology.
Eqs.(\ref{beg}) are called {\sl self-dual equations} while
eqs.(\ref{beg9}) are  the corresponding {\sl anti self-dual
equations}. Solutions to eqs.(\ref{beg}) correspond to positive
magnetic flux,  while those to (\ref{beg9}) give negative magnetic
flux. Note that  our convention coincide with that in (\cite{JMW})
and is the opposite to that in  \cite{BLP}, where, in our
terminology,   anti self-dual solutions are discussed in detail
and  a critical value of the noncommutative parameter is found,
$\theta_c=1/\eta^2$, such that solutions cease to exist when
$\theta > \theta_c$. Now, as stressed above,  in the
noncommutative case, the presence of the parity breaking $\theta$
parameter renders the connection between the anti self-dual and
the self-dual case    non-trivial, in contrast to what happens in
the commutative case where it is straightforward.

In what follows, we construct {\sl exact solutions to the
self-dual equations}
 (\ref{beg})
for arbitrary values of $\theta$ and in this sense, our
calculation complements those in \cite{JMW} and \cite{BLP}. To
this end, we propose the following ansatz
\begin{eqnarray}
\hat A_z&=&\frac{i}{\sqrt \theta} \sum_n
(\sqrt{n+1}-\sqrt{n+2}+e_n)|n+1\rangle \langle n| \label{an0}\\
\hat \phi&=& \eta \sum_n f_n |n\rangle \langle n+1| \label{an}
\end{eqnarray}
Notice that the Higgs field can be rewritten as
\begin{equation}
\hat \phi= \eta \frac{f(\hat N)}{\sqrt{\hat N+1}}
\frac{X^1+iX^2}{\sqrt{2\theta}}
\end{equation}
where $\hat N = \hat a^\dagger \hat a$
and $\langle|f(\hat N)|n\rangle=f_n$. This sould be compared with
 the ansatz in
the commutative case,
\begin{equation}
\phi= \eta \,g(|z|) \,z
\end{equation}
with $g(0)$ to be determined by solving the Bogomol'nyi
equations and requiring that at infinity $g(|z|) \sim 1/|z|$. This has
been done in \cite{dVS} with the result
\be
g(0)^2=0.72791
\label{nume}
\ee

In the same way, introducing the ansatz (\ref{an0})-(\ref{an}) we
expect to derive a recurrence relation for $f_n$
whose solution is uniquely determined by requiring that
$f(\infty)\to 1$.

Notice also that the flux-tube solutions presented in
\cite{GMS},\cite{JKL}  correspond to the choice of coefficients
$e_n=0$ and $f_n = 1$, leading to ``quasi pure gauge'' solutions
(which, in the $\theta \to 0$ limit give
 singular
vortex solutions with magnetic field $B = \delta^{(2)}(x)\,$). What we
are looking for here is to determine, through recurrence relations
deriving from (\ref{beg})-(\ref{an}), the non-trivial values for
$e_n,f_n$  that correspond to exact solutions, which should lead to the
regular ones found in \cite{dVS} in the commutative $\theta\to 0$ case.
 In fact, this ansatz can be seen as the analogous to
  performing, in the commutative case, a $U(1)$ singular gauge transformation
 $ \exp(in\varphi)$ on $|\phi(r)|$; the condition $|\phi(0)| = 0$
 ensures the regularity of the solution. In noncommutative space, the equivalent of
  such a procedure is to apply an operator
$S^n$ with $\hat S$ the shift operator defined as \cite{GMS}
\begin{equation}
\hat S = \sum_{k}|k\rangle \langle k+1 | \ee
Ansatz (\ref{an}) just corresponds to a combination of bra and kets like in $S$
but with arbitrary coefficients $f_n$. It is easy to also see that
the compatible ansatz for the gauge field is just (\ref{an0}).

Now, in order to determine the up to now arbitrary coefficients
$f_n,e_n$, we  plug  ansatz (\ref{an0})-(\ref{an}) in
eqs.(\ref{beg}) getting the following recurrence relations
\begin{eqnarray}
\sqrt{(n+2)}(f_{n+1}-f_{n})-e_n f_{n+1}&=&0  \nonumber\\
2\sqrt{(n+1)} e_{n-1}-e^2_{n-1}- 2\sqrt{(n+2)} e_{n}+e^2_{n}&=&-
\theta \eta^2 (f_n^2-1) \label{ran}
\end{eqnarray}
  This coupled system
 can be combined to give for $f_n$
\begin{eqnarray}
f^2_{1} &=& \frac{2 f^2_0}{1+\theta \eta^2 - \theta \eta^2
(f_0^2)} \nonumber\\ f^2_{n+1} &=& \frac{(n+2) f^4_n}{f_n^2
-\theta \eta^2 f_n^2(f_n^2-1)+(n+1)f^2_{n-1}} \, \;\;\;\; n >0
\label{rec}
\end{eqnarray}
Given a value for $f_0$  one can then determine all $f_n$'s from
(\ref{rec}). The correct value  for $f_0$  should make $f_n^2 \to
1$ asymptotically so that boundary conditions are satisfied. The
values of these coefficients will depend on the choice of the
dimensionless parameter $\theta \eta^2$.
For small $\theta$ we have checked that we re-obtain the values
for the commutative
solution. Indeed,
\begin{equation}
\frac{f_0^2}{2\eta^2\theta} =0.72792  \,\,\,\,\,\,\,\,\, \theta<<1
\end{equation}
(compare with eq.(\ref{nume})), while for large $\theta$ we reobtain the result of ref. \cite{JMW}
\begin{equation}
f_0^2=1-\frac{1}{\eta^2\theta} \,\,\,\,\,\,\,\,\, \theta>>1
\end{equation}
Exploring the whole range of $\theta \eta^2$, one finds that
the vortex solution with $+1$ units of magnetic flux  exists  in
all the intermediate range.
 As an example,
we list three representative values,
\begin{eqnarray}
\theta \eta^2 = 0.5, & & f_0^2 = 0.40069\ldots \nonumber\\
\theta \eta^2 = 1.0, & & f_0^2 = 0.56029\ldots \nonumber\\
\theta \eta^2 = 2.0, & & f_0^2 = 0.70670\ldots
\end{eqnarray}
Once all $f_n's$ and $e_n's$ are calculated, one can compute the
magnetic field, using for example the formula
\be
\hat B = \eta^2 \sum_{n=0}^{\infty} \left(1 - f_n^2\right) \vert n
\rangle \langle n \vert \label{ayu} \ee
or, using the explicit formula for $\vert n \rangle \langle  n
\vert $ in configuration space \cite{GMS}
\be
B(r) = 2 \eta^2 \sum_{n=0}^{\infty}(-1)^n \left(1 - f_n^2\right)
\exp(-\frac{r^2}{\theta}) L_n(2\frac{r^2}{\theta}) \label{B} \ee
where $L_n$ are the Laguerre polynomials.

We show in figure 1 the resulting magnetic field $B$ as a function
of $r\eta$. For $\theta = 0$ one recovers the result for self dual
Nielsen-Olesen vortices in ordinary space \cite{dVS}. As $\theta$
grows, the maximum for B decreases and the vortex is less
localized with total area such that the magnetic flux remains
equal to 1. It is important to stress that we have found
noncommutative self-dual vortex solutions in the whole range of
$\theta$, in agreement with the analysis for large and small
$\theta$ presented in \cite{JMW}.

As $\theta$ becomes larger, one needs more and more precision in
order to match the value of $f_0$ so that the vortex has the
adequate behavior at infinity, but a solution can be always found
(this should be contrasted with the anti self-dual case discussed
in \cite{BLP}). One can easily integrate $B(r)$ in (\ref{B}) and
check that the magnetic flux $\Phi$, which can also be written as
\be
\Phi = 2 \pi \theta {\rm Tr} \hat B
\label{flujo}
\ee
 gives, for the exact solution,
\be
\frac{\Phi}{2\pi} = 1 \ee
We have also computed the energy by inserting our vortex solution
directly in eq.(\ref{enb}). As expected, the solution saturates
the bound giving
\be
E = 2\pi \eta^2 \ee
In summary, we have constructed exact regular vortex solutions to
the self-dual equations of Abelian Higgs model in  noncommutative
space for arbitrary values of $\theta$. The solution corresponds
to  a magnetic flux $\Phi=2\pi$ (solutions with $\Phi = 2\pi n$
can be constructed by  straightforward generalization of our
procedure). For $\theta \to 0$ it converges to the commutative
Nielsen-Olesen solution while, for growing $\theta $ the flux tube
becomes more and more delocalized. The connection between
self-dual and anti-self-dual solutions deserves a thorough
investigation which we hope to present elsewhere.

\vspace{1 cm}

\noindent\underline{Acknowledgements}: This work is partially
supported by CICBA, CONICET (PIP 4330/96), ANPCYT (PICT 97/2285).
G.S.L.  and E.F.M. are partially supported by Fundaci\'on
Antorchas.



\begin{figure}
\centerline{
\psfig{figure=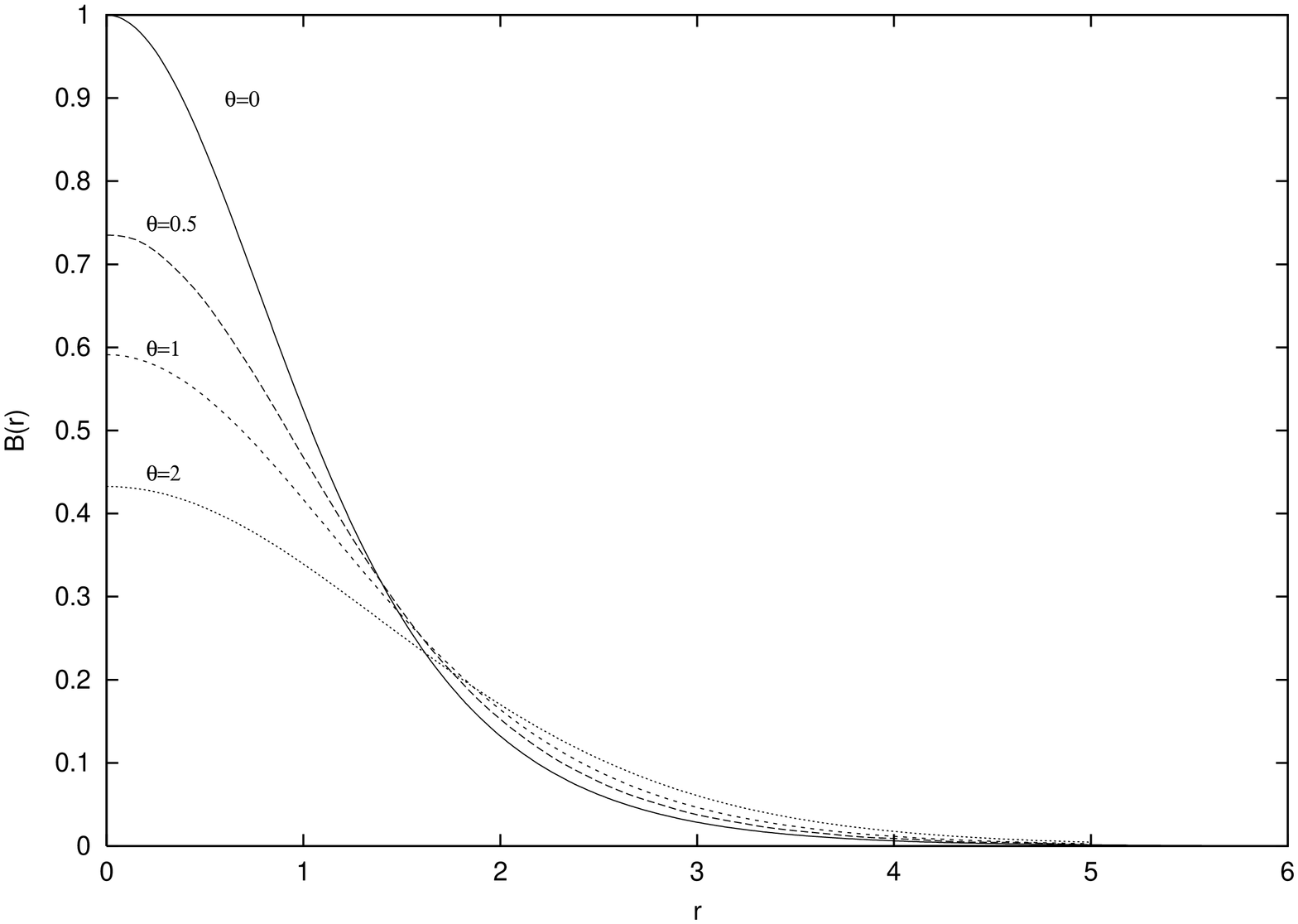,height=18cm,width=18cm,angle=-90}}
\smallskip
\caption{ Magnetic field of the vortex as a function of the radial
coordinate (in units of $\eta$) for different values of the
anticommuting parameter $\theta$ (in units of $\eta^2$). The curve
for $\theta = 0$ coincides with that of the ordinary
Nielsen-Olesen vortex.
\label{fig1} }
\end{figure}
\vspace{-2 cm}

\end{document}